# Exact Discrete Stochastic Simulation with Deep-Learning-Scale Gradient Optimization

Jose M. G. Vilar[1,2,*] and Leonor Saiz[3,*]

[1] Biofisika Institute (CSIC, UPV/EHU), University of the Basque Country (UPV/EHU), P.O. Box 644, 48080 Bilbao, Spain

[2] IKERBASQUE, Basque Foundation for Science, 48011 Bilbao, Spain

[3] Department of Biomedical Engineering, University of California, 451 E. Health Sciences Drive, Davis, CA 95616, USA

[*]Correspondence to: j.vilar@ikerbasque.org or lsaiz@ucdavis.edu

## Abstract

Exact stochastic simulation of continuous-time Markov chains (CTMCs) is essential when discreteness and noise drive system behavior, but the hard categorical event selection in Gillespie-type algorithms blocks gradient-based learning. We eliminate this constraint by decoupling forward simulation from backward differentiation, with hard categorical sampling generating exact trajectories and gradients propagating through a continuous massively-parallel Gumbel–Softmax straight-through surrogate. Our approach enables accurate optimization at parameter scales over four orders of magnitude beyond existing simulators. We validate for accuracy, scalability, and reliability on a reversible dimerization model (0.09% error), a genetic oscillator (1.2% error), a 203,796-parameter gene regulatory network achieving 98.4% MNIST accuracy (a prototypical deep-learning multilayer perceptron benchmark), and experimental patch-clamp recordings of ion channel gating ($R^2 = 0.987$) in the single-channel regime. Our GPU implementation delivers 1.9 billion steps per second, matching the scale of non-differentiable simulators. By making exact stochastic simulation massively parallel and autodiff-compatible, our results enable high-dimensional parameter inference and inverse design across systems biology, chemical kinetics, physics, and related CTMC-governed domains.

## 1. Introduction

Stochastic fluctuations are not merely noise; in systems ranging from gene regulatory networks to viral epidemiology and nucleation kinetics, they are the drivers of phenotypic variability, extinction events, and symmetry breaking (*1-12*). To model these phenomena, exact discrete-event simulation algorithms, such as the Gillespie algorithm (*13*) and the Bortz-Kalos-Lebowitz (BKL) method (*14*), serve as the gold standard, providing mathematically rigorous trajectories of the underlying CTMC (*3, 15*). However, this physical rigor has come at a steep computational cost because the discrete, non-differentiable nature of event selection breaks the computational graph, rendering exact stochastic simulation fundamentally incompatible with the gradient-based optimization techniques that have revolutionized deep learning (*16*).

Consequently, parameter inference and design for stochastic models have remained historically bottlenecked by the curse of dimensionality (*17, 18*). For decades, the field has been relegated to likelihood-free methods, such as approximate Bayesian computation (ABC), which do not require gradients but scale poorly, limiting inference to models with fewer than a dozen parameters (*3, 19, 20*). Alternative gradient-estimation techniques exist but suffer from critical weaknesses in this context. Likelihood-ratio (score-function) estimators (*21*) provide unbiased gradients but exhibit variance that explodes with trajectory length, making them impractical for large systems (*22*) and often requiring clever system-specific approaches (*23, 24*).

Other sophisticated unbiased approaches, such as the Poisson Path Algorithm (PPA) (*25*), significantly reduce this variance by using an integral representation of the sensitivity. While these methods, along with common random number (CRN) finite-difference schemes (*26*), make gradient computation feasible for optimization in lower-dimensional spaces, they are inherently limited by their scaling. The computational cost grows linearly with the number of parameters because these estimators must be evaluated for each parameter individually.

This linear-scaling bottleneck effectively prevents their application to the high-dimensional parameter spaces required for complex tasks like neural network-style systems. Conversely, prior soft-forward methods achieve differentiability by directly approximating the dynamics (e.g., continuous reaction mixtures), but this introduces a simulation-reality mismatch where the optimized model no longer represents the discrete stochastic physics, and have thus far been demonstrated only in low-dimensional systems (*27*).

We show that physical exactness and differentiability are not mutually exclusive at deep learning scales. Our central insight is the complete decoupling of the forward simulation from the backward differentiation. In the forward pass, we retain standard, hard categorical sampling, ensuring that the solver captures the true intrinsic noise and discreteness of the system without approximation. In the backward pass, we bridge the discontinuity using a massively-parallel Gumbel-Softmax straight-through estimator (*22, 28, 29*), a well-stablished technique in deep learning for discrete (categorical) choices. This surrogate gradient is known to allow the optimizer to see through discrete events, guiding parameters via a tunable signal while the forward dynamics remain statistically exact (*22, 28, 29*).

Our framework removes the dimensionality barrier entirely, enabling us to scale from handfuls of parameters to hundreds of thousands. We validate the approach across a hierarchy of complexity spanning five orders of magnitude, beginning with precision benchmarks where we recover parameters for a reversible dimerization system with 0.09% error and successfully learn the specific rates required to sustain limit-cycle oscillations in a genetic oscillator (*30*), often used as an exemple of difficult parameter identifiability (*31*), with 1.2% error. To demonstrate extreme scalability, we train a 203,796-parameter gene regulatory network to classify MNIST digits (*32*) with 98.4% accuracy, matching standard multilayer perceptron benchmarks (*33*) and proving that stochastic reaction networks can be optimized for complex computation. Finally, we validate the method on experimental data by inferring ion channel gating kinetics from patch-clamp recordings (*34*) ($R^2 = 0.987$). This latter application is particularly significant because with only two channels in the system, the dynamics are dominated by discrete transitions with no law of large numbers smoothing, confirming that our method genuinely optimizes inherently discrete stochastic processes.

By decoupling physical exactness in the forward pass from stable, massively parallel gradient computation in the backward pass, we establish exact stochastic simulation as a scalable, differentiable primitive supported by a GPU implementation capable of 1.9 billion steps per second. This removes a longstanding barrier in computational modeling, enabling high-dimensional parameter inference and inverse design across systems biology, chemistry, physics, epidemiology, and other domains governed by continuous-time Markov chains. The approach opens the door to deep-learning gradient-based optimization of realistic stochastic models and the automated engineering of reaction circuits at scales previously out of reach.

We note that concurrent and independent work by Mottes et al. (*35*) applies a similar straight-through Gumbel-Softmax strategy to stochastic kinetic models and validates the method on simulated telegraph promoter gene expression and stochastic thermodynamics problems in the low-dimensional regime. While the methods share the same core forward/backward decoupling principle, our work is specifically focused on overcoming the dimensionality barrier in exact stochastic optimization. By scaling to parameter spaces over four orders of magnitude larger, our framework enables deep-learning-scale tasks, including a 203,796-parameter gene regulatory network for MNIST classification via high-throughput GPU implementation, while remaining highly robust to accurately optimize directly on experimental data (patch-clamp recordings in the extreme discrete regime of *N=2* ion channels).

## 2. The Gillespie Stochastic Simulation Algorithm

We develop our framework in the context of the Gillespie algorithm, the canonical method for exact CTMC simulation in chemical kinetics, though the approach generalizes to any system of competing Poisson processes. Consider a well-mixed chemical system containing $N$ molecular species whose copy numbers at time $t$ are represented by the state vector $\mathbf{X}(t) = (X_1(t), \ldots, X_N(t))^\top$. The system evolves through $M$ reaction channels, where each reaction $R_j$ is characterized by a propensity function $a_j(\mathbf{X}; \boldsymbol{\theta})$ and a stoichiometry vector $\mathbf{v}_j \in \mathbb{Z}^N$. The propensity function gives the probability per unit time that reaction $R_j$ fires given the current state and depends on kinetic parameters $\boldsymbol{\theta}$ that we wish to infer or optimize. When reaction $R_j$ occurs after a time $\tau$, the state updates according to $\mathbf{X}(t + \tau) = \mathbf{X}(t) + \mathbf{v}_j$.

The state vector $\mathbf{X}(t)$ evolves as a continuous-time Markov chain whose dynamics are governed by the chemical master equation. Let $P(\mathbf{x}, t)$ denote the probability of finding the system in state $\mathbf{x}$ at time $t$. The master equation takes the form

$$\frac{\partial P(\mathbf{x}, t)}{\partial t} = \sum_{j=1}^{M} \left[ a_j(\mathbf{x} - \mathbf{v}_j; \boldsymbol{\theta}) P(\mathbf{x} - \mathbf{v}_j, t) - a_j(\mathbf{x}; \boldsymbol{\theta}) P(\mathbf{x}, t) \right]$$

(1)

where the first term represents probability flux into state $\mathbf{x}$ from states that can reach it via reaction $j$, and the second term represents flux out of state $\mathbf{x}$ due to reaction $j$ firing (*15*). Analytical solutions to the master equation exist only for simple systems, motivating the use of stochastic simulation to sample from the distribution $P(\mathbf{x}, t)$.

The Gillespie algorithm generates exact sample paths from the master equation through the following procedure. Given the current state $\mathbf{X}$ at time $t$, first compute the propensity $a_j(\mathbf{X}; \boldsymbol{\theta})$ for each reaction and the total propensity $a_0 = \sum_{j=1}^{M} a_j(\mathbf{X}; \boldsymbol{\theta})$. Sample the waiting time until the next reaction from an exponential distribution with rate $a_0$, which can be accomplished via the inverse transform $\tau = -\ln(u_1)/a_0$ where $u_1 \sim \text{Uniform}(0,1)$. Select which reaction fires by sampling from the categorical distribution with probabilities $\pi_j = a_j/a_0$. Update the state to $\mathbf{X} + \mathbf{v}_j$ and advance time to $t + \tau$. This procedure repeats until a terminal condition is reached.

The propensity functions encode the kinetics of each reaction and typically follow mass-action form. A first-order reaction $S_i \rightarrow \cdots$ has propensity $a = kX_i$ where $k$ is the rate constant. A second-order reaction $S_i + S_j \rightarrow \cdots$ with $i \neq j$ has propensity $a = kX_iX_j$. A homodimerization reaction $2S_i \rightarrow \cdots$ has propensity $a = kX_i(X_i - 1)/2$, accounting for the number of distinct pairs. More complex propensity functions arise in enzyme kinetics and cooperative binding, but the Gillespie algorithm applies unchanged regardless of the functional form.

The computational cost of the Gillespie algorithm scales with the number of reaction events, which in turn depends on the total propensity and simulation duration. For systems with fast reactions or large molecular populations, direct simulation can become prohibitively expensive. Various approximation methods trade exactness for computational efficiency, including tau-leaping, the chemical Langevin equation, and more recent mapping-based approaches such as Holimap (*36-38*). Here, we focus on exact simulation and address computational challenges through massive parallelization rather than approximation.

## 3. Gumbel-Softmax Reparameterization for Differentiable Simulation

### 3.1 The Differentiability Challenge

The fundamental obstacle to gradient-based optimization through stochastic simulation is the discrete nature of reaction selection. At each step, the Gillespie algorithm samples a reaction index $j$ from the categorical distribution with probabilities proportional to propensities. This sampling operation is not differentiable; namely, the gradient of a discrete sample with respect to its underlying probabilities is zero almost everywhere and undefined at the sampling boundaries. Standard automatic differentiation frameworks cannot propagate gradients through such operations.

The waiting time presents a lesser challenge because it is a continuous random variable. Given the total propensity $a_0$, the waiting time $\tau \sim \text{Exponential}(a_0)$ can be reparameterized as $\tau = -\ln(u)/a_0$ where $u \sim \text{Uniform}(0,1)$. This reparameterization expresses $\tau$ as a differentiable function of $a_0$ and hence of the underlying parameters $\boldsymbol{\theta}$, with the randomness absorbed into the fixed random variable $u$. Gradients flow through $\tau$ via the chain rule: $\partial\tau / \partial\theta = -(\tau/a_0) \cdot \partial a_0 / \partial\theta$.

The reaction selection step requires a different approach. We seek a continuous relaxation of categorical sampling that preserves essential statistical properties while enabling gradient computation. The relaxation should approach exact categorical sampling in an appropriate limit, ensuring that optimization with respect to the relaxed objective produces parameters that are valid for the exact dynamics.

### 3.2 The Gumbel-Max Reparameterization

The Gumbel-Max reparameterization provides a representation of categorical sampling that is amenable to continuous relaxation (*39*). Let $G_1, \ldots, G_M$ be independent random variables with the standard Gumbel distribution, which has probability density function $f(g) = \exp(-g - \exp(-g))$ and can be sampled via $G = -\ln(-\ln(u))$ where $u \sim \text{Uniform}(0,1)$. The Gumbel-Max theorem states that $j^* = \arg\max_{j \in \{1,\ldots,M\}} \left[ \log \pi_j + G_j \right]$ has the same distribution as a sample from the categorical distribution with probabilities $(\pi_1, \ldots, \pi_M)$. For reaction selection, we can therefore write

$$j^* = \arg\max_{j \in \{1,\ldots,M\}} \left[ \log a_j(\mathbf{X}; \boldsymbol{\theta}) + G_j \right]$$

where the normalization by total propensity is unnecessary because argmax is invariant to additive constants.

This representation transforms categorical sampling into an optimization problem, but the argmax operation remains non-differentiable. The key observation is that argmax can be approximated by softmax with a temperature parameter that controls the sharpness of the approximation.

### 3.3 The Gumbel-Softmax Relaxation

The Gumbel-Softmax relaxation replaces the hard argmax with a temperature-controlled softmax (*28*). Define the soft sample

$$\tilde{y}_j = \frac{\exp\big((\log a_j(\mathbf{X}; \boldsymbol{\theta}) + G_j)/T\big)}{\sum_{k=1}^{M} \exp\big((\log a_k(\mathbf{X}; \boldsymbol{\theta}) + G_k)/T\big)}$$

(3)

where $T > 0$ is the temperature parameter. The vector $\tilde{\mathbf{y}} = (\tilde{y}_1, \ldots, \tilde{y}_M)$ lies on the probability simplex and is a differentiable function of the propensities and hence of the parameters $\boldsymbol{\theta}$. As $T \to 0$, the softmax output approaches a one-hot vector corresponding to the argmax, recovering exact categorical sampling. As $T \to \infty$, the output approaches a uniform distribution regardless of the propensities.

The soft sample $\tilde{\mathbf{y}}$ can be used to compute a relaxed state update. Instead of applying the stoichiometry vector $\mathbf{v}_{j^*}$ corresponding to the selected reaction, we apply the expected stoichiometry

$$\Delta \mathbf{X}_{\text{soft}} = \sum_{j=1}^{M} \tilde{y}_j \, \mathbf{v}_j$$

(4)

which is a weighted average of all possible updates. This relaxed update is differentiable with respect to the parameters and can be used to simulate approximate trajectories for optimization.

However, using soft updates throughout the simulation fundamentally changes the dynamics. The relaxed state $\mathbf{X}_{\text{soft}}$ does not correspond to any realization of the exact stochastic process, as molecular copy numbers are inherently discrete. Parameters optimized to fit relaxed trajectories may not produce correct behavior under exact simulation, limiting the utility of the approach for applications where exact stochasticity matters.

### 3.4 The Straight-Through Estimator

The straight-through estimator resolves the conflict between exactness and differentiability by decoupling the forward and backward passes (Figure 1). Let $\mathbf{y} = \text{one\_hot}(j^*) \in \{0,1\}^M$ denote the hard reaction indicator from the forward pass obtained from Eq. (3). The straight-through construction

$$\mathbf{y}_{\text{ST}} = \text{stopgrad}(\mathbf{y} - \tilde{\mathbf{y}}) + \tilde{\mathbf{y}}$$

(5)

ensures that the forward computation uses the hard sample $\mathbf{y}$ while the backward computation differentiates through the soft sample $\tilde{\mathbf{y}}$. Here, stopgrad is the do-not-differentiate-through-this-part operator, present in deep learning frameworks, that makes the forward pass exact and the backward pass differentiable. Therefore, the forward pass uses hard categorical samples to generate exact trajectories:

$$\mathbf{X}'_{\text{ST}} = \mathbf{X} + \mathbf{v}_{j^*}.$$

(6)

This produces a state update that is statistically identical to standard Gillespie simulation. The trajectory consists of discrete states connected by single-reaction jumps, exactly as in the original algorithm.

For gradient computation in the backward pass, it replaces the hard argmax with a Gumbel-Softmax relaxation. Automatic differentiation then propagates gradients as if the state update were from $\Delta\mathbf{X}_{\text{soft}}$ (Eq. 4), yielding the surrogate Jacobian

$$\frac{\partial \mathbf{X}_{\text{ST}}}{\partial \boldsymbol{\theta}} = \sum_{j=1}^{M} \mathbf{v}_j \frac{\partial \tilde{y}_j}{\partial \boldsymbol{\theta}}.$$

(7)

At low temperatures, the soft samples $\tilde{\mathbf{y}}$ approach one-hot vectors, reducing bias but increasing gradient variance. At high temperatures, the soft samples become more uniform, providing smoother gradients at the cost of a less accurate approximation to the discrete dynamics. In practice, temperature annealing from higher to lower values over the course of training often improves convergence.

The straight-through estimator is biased because the gradient it computes is not the exact gradient of the expected objective with respect to parameters. However, for optimization purposes, an unbiased gradient is not required; what matters is that the gradient provides a useful descent direction. Empirical evidence across the applications presented here demonstrates that straight-through gradients are effective for optimization despite their potential bias.

### 3.5 Implementation Details

Our implementation combines several techniques to achieve computational efficiency and numerical stability. The framework is implemented in TensorFlow 2.20, which provides automatic differentiation, GPU acceleration, and XLA (Accelerated Linear Algebra) JIT compilation. The straight-through estimator is expressed entirely within TensorFlow's computational graph using `tf.stop_gradient` to decouple the forward and backward passes, enabling standard backpropagation through the surrogate without custom gradient definitions. All simulations are performed in parallel across an ensemble of independent trajectories, exploiting GPU parallelism through vectorized operations over the batch dimension. The ensemble size is a tunable hyperparameter that trades memory usage against variance in the gradient estimate.

The Gumbel-Softmax temperature $T$ is a critical hyperparameter whose optimal value is system-dependent. A key observation is that at low $T$, the Gumbel perturbations average out across the ensemble and the gradient estimates are effectively unbiased. The tradeoff at low temperature is increased variance rather than bias, which is addressed through sufficient ensemble size. Temperature schedules used here span several orders of magnitude across applications: the dimerization system uses geometric annealing from $T = 1.0$ to $T = 0.001$; the genetic oscillator uses a fixed temperature $T = 3 \times 10^{-6}$; the MNIST classification network anneals from $T = 2.0$ to $T = 0.2$; and the ion channel system anneals from $T = 0.05$ to $T = 0.0005$. In general, systems with few reactions and large ensembles tolerate very low temperatures, while high-dimensional systems benefit from warmer temperatures that provide smoother gradients.

For the high-dimensional MNIST classification task, we employ a two-phase training protocol. In the initial phase (the first 16 of 80 epochs), the forward pass uses soft Gumbel-Softmax samples rather than hard categorical draws, providing gradient signal that can traverse the flat regions of the discrete loss landscape (discrete molecular changes are too large in the earlier stages of training to decrease the loss function). In the second phase, training switches to the straight-through estimator with hard forward samples, refining parameters under the exact discrete dynamics. The remaining three systems (dimerization, genetic oscillator, and ion channel) use the straight-through estimator throughout training, as their lower dimensionality and larger relative ensemble sizes provide sufficient gradient signal without a soft warm-up phase.

The RMSprop optimizer is used for all four systems, with learning rates coupled to the temperature annealing schedule. Learning rates are reduced as training progresses, either through explicit annealing schedules or through coupling to the temperature parameter. For the dimerization and oscillator systems, learning rate annealing follows a geometric schedule. For the ion channel system, the learning rate is proportional to the current temperature. The MNIST system uses a fixed learning rate of $10^{-2}$ with stochastic weight averaging applied over the final epochs to improve generalization.

The loss function for parameter inference is the mean squared error between simulated and target trajectory statistics. For the dimerization and ion channel systems, we match ensemble-averaged trajectories at each time point. For the oscillator, the target data is generated via a start-point map, in which short trajectory segments are simulated from start points sampled along a reference trajectory, and the loss measures deviation between simulated and target segments. For MNIST classification, the loss is the categorical cross-entropy between softmax-transformed output gene copy numbers and one-hot encoded digit labels.

## 4. Results

### 4.1 Reversible Dimerization

We first validate the framework on a prototypical reversible dimerization reaction system (Figure 2A and Supplementary Note 1):

$$A + B \underset{k_2}{\overset{k_1}{\rightleftharpoons}} C$$

(8)

This system involves three molecular species and two reactions with rate constants $k_1$ (forward, dimerization) and $k_2$ (reverse, dissociation). The propensity functions follow mass-action kinetics: $a_1(\mathbf{X}) = k_1 X_A X_B$ for the forward reaction and $a_2(\mathbf{X}) = k_2 X_C$ for the reverse reaction (15). The stoichiometric vectors are $\mathbf{v}_1 = (-1, -1, +1)^\top$ and $\mathbf{v}_2 = (+1, +1, -1)^\top$.

We generated synthetic trajectory data using ground truth parameters $k_1 = 0.01$ and $k_2 = 0.32$, with initial conditions $X_A(0) = 100$, $X_B(0) = 90$, and $X_C(0) = 0$. The target data consisted of ensemble-averaged trajectories computed from 100,000 independent simulations, providing low-variance estimates of the expected molecular counts over time. For parameter inference, we used ensembles of 100,000 parallel trajectories per gradient computation, simulating each trajectory for 250 steps to observe both transient and near-equilibrium behavior. The loss function is the mean squared error (MSE) between the simulated and target ensemble-averaged trajectories for all three species (A, B, and C). Because the stochastic Gillespie trajectories are defined on irregular, event-driven time grids, each simulated trajectory is first linearly interpolated onto a uniform target time grid before averaging across the ensemble. The MSE is then computed at each grid point and averaged over time, comparing the model mean to the target mean for all three species simultaneously.

Parameter optimization proceeded over 250 epochs using the RMSprop optimizer. Parameters were represented in log-space to ensure positivity: the optimization variables were $\theta_1 = \log k_1$ and $\theta_2 = \log k_2$. Initial parameter guesses were set arbitrarily at $k_1 = 0.125$ and $k_2 = 0.025$.

The optimization converged rapidly, with the loss decreasing by three orders of magnitude within the first 100 epochs (Figure 2C). Across multiple inference runs with varying ground truth parameters, the learned parameters achieved mean absolute percentage errors of 0.06% for $k_1$ (95% CI: 0.04%–0.08%) and 0.13% for $k_2$ (95% CI: 0.07%–0.18%). This level of accuracy exceeds the requirements of most practical applications.

To assess robustness across different kinetic regimes, we repeated the inference procedure for a range of reverse rate constants $k_2 \in \{0.01, 0.02, 0.04, 0.08, 0.16, 0.32, 0.64, 1.28\}$ while holding $k_1 = 0.01$ fixed. Each condition used 100,000 target trajectories and 100,000 model simulations per iteration (250 steps, 250 epochs). This spans equilibrium constants $K = k_1/k_2$ from 0.0078 to 1.0, corresponding to strongly product-favored through balanced conditions (Figure 2D–E). The overall average mean absolute percentage error was 0.09% (95% CI: 0.06%–0.13%), demonstrating that the method succeeds across diverse kinetic regimes.

## 4.2 Genetic Oscillator

The genetic oscillator by Vilar et al. (*30*), which involves complex nonlinear dynamics, multiple time scales, and emergent oscillatory behavior (Figure 3A and Supplementary Note 2), provides a more challenging test case and is frequently employed as an exemplar of difficult parameter identifiability problems (*31*). This system, originally proposed as a model of circadian rhythm generation, consists of two genes encoding an activator protein and a repressor protein. The activator promotes its own expression through positive feedback and activates repressor expression through a feed-forward motif. The repressor inhibits activator function through direct binding, creating the delayed negative feedback necessary for sustained oscillations.

The complete reaction network involves nine molecular species: the activator gene in inactive ($D_A$) and active ($D_A^{'}$) states, the repressor gene in inactive ($D_R$) and active ($D_R^{'}$) states, activator mRNA ($M_A$), repressor mRNA ($M_R$), activator protein ($A$), repressor protein ($R$), and the activator-repressor complex ($C$). The network comprises sixteen reactions including gene activation and deactivation, transcription, translation, degradation, and complex formation.

We focused on inferring five protein parameters that control the oscillatory dynamics: the activator translation rate $\beta_A = 50.0$, the repressor translation rate $\beta_R = 5.0$, the activator protein degradation rate $\delta_A = 1.0$, the repressor protein degradation rate $\delta_R = 0.2$, and the complex formation rate $\gamma_C = 2.0$. The remaining eleven parameters were held fixed at their nominal values.

This partial inference scenario reflects realistic situations where some parameters are known from biochemical measurements while others must be inferred from dynamic data.

Synthetic target data was generated using a start-point map approach. A reference trajectory of 2,400,000 steps was simulated with ground truth parameters, from which 262,144 start points were sampled. Each training iteration simulated short trajectory segments of horizon $T_{segment} = 150$ steps from these start points, using 25 replicate simulations per start point and batches of 8,192 start points per gradient step (204,800 total simulations per step). The loss function measured the discrepancy (mean squared error) between the rate of change of simulated and target trajectory segments of the three protein species (activator A, repressor R, and complex C). Each rate is estimated as the change in mean species count from the initial state to the mean of the final 100 simulation steps, divided by the corresponding elapsed time.

Parameter inference used the RMSprop optimizer with an initial learning rate of $10^{-1}$ and a geometric decay schedule. Training proceeded for 3,000 epochs. Gumbel-Softmax temperature was fixed at $T = 3 \times 10^{-6}$. Total training time was approximately 19 minutes per run on a single GPU. To ensure robustness, we avoided aggressive annealing and instead employed a median-based Polyak-Ruppert averaging scheme (*40*) over the final 25% of epochs and pooling across runs.

The inferred parameters achieved excellent agreement with ground truth values (Figure 3C–D). Across ten independent inference runs, the global median parameter estimates were: $\hat{\beta}_A = 49.26$ (95% CI: 48.39–50.13), $\hat{\beta}_R = 5.03$ (95% CI: 4.81–5.25), $\hat{\delta}_A = 0.99$ (95% CI: 0.97–1.01), $\hat{\delta}_R = 0.204$ (95% CI: 0.196–0.211), and $\hat{\gamma}_C = 2.01$ (95% CI: 1.92–2.10). The mean absolute percentage error (MAPE) across all five parameters was 1.23% (95% CI: 0.93%–1.71%), with per-run instantaneous MAPE trajectories in Figure 3F and the histogram of bootstrapped MAPE values in Figure 3B.

Validation of the inferred parameters extended beyond point estimates to verification of emergent dynamics (Figure 3E). Analysis comparing trajectories generated with learned versus true parameters showed virtually identical results (same random number sequence was used in both cases). The oscillation period, amplitude, and waveform shape were all faithfully reproduced. This agreement confirms that the optimization successfully captured the functional behavior of the system, not merely individual parameter values.

### 4.3 Gene Regulatory Network for Image Classification

To demonstrate that the framework scales to high-dimensional parameter spaces where gradient-based optimization becomes not merely helpful but essential, we constructed a gene regulatory network capable of classifying handwritten digits from the MNIST dataset (Supplementary Note 3) (*32*). This benchmark is not intended to outcompete neural networks on accuracy, but it is designed to establish that exact stochastic simulation can now be optimized competitively at scales previously considered intractable. With 203,796 trainable parameters, this system is four orders of magnitude larger than any previously demonstrated differentiable stochastic simulation, entering a regime where derivative-free methods such as grid search or evolutionary algorithms are computationally infeasible.

The network architecture draws inspiration from biological gene regulatory systems (*41*) while incorporating structure necessary for classification (Figure 4A). The input layer consists of 784 transcription factor species whose concentrations encode the pixel intensities of a 28×28 grayscale image. These transcription factors regulate a hidden layer of 256 genes. The hidden genes in turn regulate 10 output genes corresponding to the digit classes 0 through 9. Each gene has a promoter with two coarse-grained states (OFF and ON). The probability of the ON state is derived from a Boltzmann weight, yielding a logistic sigmoid activation function where the dimensionless activation energy operates as an affine function of the copy number of the regulating transcription factors (Supplementary Note 3). Classification proceeds by simulating the network dynamics until the output gene concentrations reach a steady state, then selecting the class with highest expression.

Each regulatory connection is parameterized by weight and bias terms that control the transcription rate of the target gene as a function of regulator copy numbers. The network contains 203,796 trainable parameters in total: $784 \times 256 = 200,704$ weights from inputs to hidden genes, 256 biases for hidden genes, $256 \times 10 = 2,560$ weights from hidden to output genes, 10 biases for output genes, and 266 degradation rate parameters. This parameter count is comparable to a medium-sized multilayer perceptron but with fundamentally different dynamics, in which the network state evolves through discrete stochastic reactions rather than deterministic matrix operations.

The network was trained on the standard MNIST training set of 60,000 images using minibatches of approximately 682 images and the RMSprop optimizer with learning rate 0.01. Each forward pass simulated 2,880 reaction steps per trajectory, allowing sufficient time for the output gene concentrations to stabilize. Training proceeded for 80 epochs, requiring approximately 4 hours on a single GPU. The final model averages the weights over the last 16 epochs of training. To control molecular copy numbers, we imposed a minimum degradation rate of 0.25 for all genes, which sets an upper bound on the achievable steady-state expression. The loss is categorical cross-entropy between the true one-hot digit label and the predicted class probabilities. The predicted probabilities are obtained by applying a softmax to the output gene copy numbers at the end of the stochastic simulation, treating the final expression level of each of the 10 output genes as a logit for the corresponding digit class.

The trained network achieved 97.8% accuracy on the held-out test set of 10,000 images using single-pass evaluation. When combining Monte Carlo averaging over 32 stochastic realizations and temporal averaging of output gene expression, accuracy improved to 98.4% (Figure 4E). To test the system's robustness at lower copy numbers, we increased the minimum degradation rate to 1.0. While this high-degradation regime initially reduced the single-pass accuracy to 84.6%, applying Monte Carlo or temporal averaging successfully restored the accuracy to over 97.0% (Figure 4F). These results demonstrate that gene regulatory network dynamics can implement nontrivial computations when parameters are appropriately tuned through gradient descent, an optimization that would be impossible without the differentiable framework developed here.

Analysis of the trained network revealed interpretable structure in the learned parameters. The input-to-hidden weights exhibited spatial organization reminiscent of Gabor-like filters, with different hidden genes responding to edges at different orientations and positions. The hidden-to-output weights showed class-specific patterns indicating which hidden gene activations support each digit classification. The gene expression dynamics during classification exhibited characteristic trajectories in which competition between output genes is resolved through the stochastic regulatory interactions (Figure 5 and Supplementary Figure 1).

## 4.4 Application to Experimental Data: Ion Channel Gating Kinetics

The preceding benchmarks demonstrate scalability on systems with known ground-truth parameters. A final critical validation is performance on experimental data, where measurement noise, model mismatch, and biological variability introduce challenges absent from synthetic benchmarks. We applied our framework to infer gating kinetics from single-channel patch-clamp recordings (*34*), a classic problem in ion channel biophysics where stochastic transitions between discrete conformational states produce the observed current fluctuations (Supplementary Note 4).

This application tests the method in three important ways beyond the synthetic benchmarks. First, it confronts experimental noise and potential model mismatch. Second, it demonstrates generality beyond chemical reactions. Ion channel conformational dynamics are governed by the same continuous-time Markov chain formalism but represent a physically distinct class of systems involving conformational changes of a single macromolecule. Third, and most importantly, it probes the regime of extreme discreteness where the quasi-continuous approximations underlying many numerical methods break down entirely. With only N = 2 channels in the patch, the observable can take only three values (0, 1, or 2 open channels). Every single stochastic transition causes a macroscopic change in the system state.

We analyzed whole-cell patch-clamp recordings from HEK293 cells expressing a voltage-gated ion channel, obtained at +40 mV holding potential (*34*). The dataset comprises 100 independent sweeps, each containing 801 timepoints sampled at 0.01 ms resolution over an 8 ms window following depolarization. Current traces were idealized to discrete channel states, yielding the number of open channels (0, 1, or 2) at each timepoint. The ensemble-averaged open probability exhibits the characteristic activation-inactivation kinetics typical of voltage-gated channels: rapid activation to a peak open probability of approximately 0.62 within 1 ms, followed by slower decay toward zero as channels enter an inactivated state (Figure 6B).

We modeled the channel as a three-state system with closed (C), open (O), and inactivated (I) conformations (*42*):

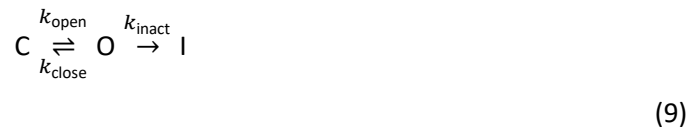

$$(9)$$

The inactivated state is absorbing on the timescale of the recording, reflecting the experimental observation that channels do not recover from inactivation during individual sweeps. This model has three unknown rate constants ($k_{open}$, $k_{close}$, $k_{inact}$), three molecular species, and three reaction channels. Each simulation begins with all channels in the closed state, matching the experimental protocol where channels recover during the inter-sweep interval at hyperpolarized holding potential.

We minimized the mean squared error between the simulated and experimental ensemble-averaged open channel counts. Explicitly, the loss is an MSE between the simulated and experimental mean open-channel counts, computed over time-binned intervals defined by the experimental recording grid. Rather than interpolating trajectories onto a grid, the simulated piecewise-constant (SSA) trajectories are exactly time-averaged over each experimental bin by computing the cumulative integral of the step function and dividing by the bin width. The experimental target for each bin is approximated as the average of the measured values at the bin endpoints. The loss is then the mean squared difference between the simulated and experimental bin-averaged open-channel counts across all time bins. Each gradient step used 262,144 parallel stochastic simulations of 20 Gillespie steps each, with trajectories interpolated onto the experimental time grid for comparison. Training proceeded for 400 epochs using the RMSprop optimizer with an initial learning rate of 0.05, annealed proportionally to the Gumbel-Softmax temperature from 0.05 down to 0.0005. Total training time was approximately 112 seconds on a single GPU.

The learned rate constants were $k_{\text{open}} = 0.750 \text{ ms}^{-1}$, $k_{\text{close}} = 0.103 \text{ ms}^{-1}$, and $k_{\text{inact}} = 1.159$ $\text{ms}^{-1}$. These values are physically reasonable: the opening rate exceeds the closing rate by a factor of seven, consistent with the high peak open probability observed experimentally, while the inactivation rate is fastest, explaining the rapid decay from peak. The ratio $k_{\text{open}}/(k_{\text{open}} + k_{\text{close}}) \approx 0.88$ predicts an equilibrium open probability in the absence of inactivation that matches the observed peak.

Validation using 30,000 independent exact Gillespie simulations with the learned parameters yielded excellent agreement with experimental data (Figures 6B, 6C, and 6E). The model mean closely tracks the experimental mean throughout the activation and inactivation phases, achieving $R^2 = 0.987$, RMSE = 0.021 open channels, and normalized RMSE of 3.5% relative to the data range. The model ensemble also captures the trial-to-trial variability observed in experimental sweeps, with individual simulated trajectories exhibiting stochastic fluctuations qualitatively similar to the experimental recordings.

Parameter trajectories during training show rapid initial convergence followed by refinement (Figure 6D). The inactivation rate $k_{\text{inact}}$ converges fastest, as it dominates the long-time behavior where all trajectories approach the absorbing state. The opening and closing rates require more iterations to disentangle, as they jointly determine the transient peak dynamics. Loss decreased by over an order of magnitude during training, with the final loss corresponding to sub-percent-level deviations between model and data.

This application demonstrates that the framework performs well on experimental data with its inherent noise and potential model mismatch. The inferred rate constants are consistent with known ion channel biophysics, and the model reproduces both the mean kinetics and the stochastic variability of the experimental recordings. Crucially, the success in this extreme low-copy-number regime, where the system can occupy only three discrete states and every molecular event produces a macroscopic observable change, demonstrates that the straight-through estimator provides valid gradients even when there is no quasi-continuous limit to fall back on. The method genuinely handles discrete stochastic dynamics.

### 4.5. Computational Performance

The computational efficiency of our implementation enables practical application to systems biology problems that would otherwise require weeks of computation. All benchmarks were performed on a single NVIDIA RTX 6000 Ada Generation GPU with 49GB memory using TensorFlow 2.20. The parallel architecture exploits the independence of trajectory realizations to achieve near-linear scaling with ensemble size up to hardware memory limits.

We benchmarked performance on the genetic oscillator system (*30*), measuring throughput as a function of ensemble size on an NVIDIA RTX 6000 Ada Generation GPU (Supplementary Figure 2 and Supplementary Note 5). For small ensembles of 100 trajectories, the implementation achieved approximately 4 million simulation steps per second. Throughput increased with ensemble size as parallel resources became fully utilized, reaching 1.37–2.30 billion steps per second, depending on the simulator variant, for ensembles of 100,000 trajectories. For comparison, the state of the art existing CPU-based SSA implementations applied to the same oscillator with the same parameters, including GillesPy2 C++ (*43*) and StochKit2 (*44*), reach 1.3–1.5 million steps per second per trajectory (*43*). Our GPU implementation achieves a 1,000-fold improvement over the fastest single-trajectory

CPU implementations when simulating large ensembles. This massive parallelism is essential for gradient estimation, which requires averaging over large ensembles to reduce variance.

The computational overhead of maintaining differentiability is minimal. Benchmarks comparing standard Gillespie, Gumbel-Max, and Gumbel-Softmax show near-identical throughput across all ensemble sizes. This demonstrates that the differentiable framework adds negligible overhead compared to non-differentiable implementations while enabling gradient-based optimization that converges in far fewer iterations than derivative-free alternatives.

Training time for parameter inference depends on the system complexity and the desired accuracy. The dimerization system converges within minutes, the genetic oscillator within tens of minutes, and the MNIST network within hours. These timescales are competitive with or faster than alternative approaches while providing the additional benefit of gradient information that can be used for sensitivity analysis, uncertainty quantification, and further optimization.

## 5. Discussion

Our results demonstrate that statistically exact stochastic simulation and deep-learning-scale gradient optimization are no longer mutually exclusive. Across benchmarks spanning five orders of magnitude, from a two-parameter kinetic system to a 203,796-parameter gene regulatory network, we show that the dimensionality barrier imposed by discrete reaction events can be overcome without relaxing the underlying physics. The central conceptual advance is the rigorous separation of roles, in which the forward pass maintains the hard categorical sampling of the underlying CTMC, while the backward pass propagates gradients via a Gumbel-Softmax straight-through estimator.

This decoupling resolves a fundamental tradeoff that has constrained the field. While the computation of gradients in discrete stochastic systems is mathematically achievable through CRN finite-difference schemes (26) or unbiased sensitivity estimators like PPA (25), these approaches are constrained by an inherent linear scaling bottleneck, effectively precluding their application to the massive models required for complex information processing. Prior differentiable approaches, such as the soft-forward Gillespie algorithm (27), achieve differentiability by replacing discrete jumps with continuous reaction mixtures. While this permits exact gradient calculation, it changes the dynamics, creating a simulation-reality mismatch that widens with trajectory length. In our approach, the simulation remains physically exact, and approximation is confined to the gradient estimator. By exposing the optimizer to the system's true intrinsic noise through the Gumbel-Max construction, while reusing the same Gumbel noise samples for the Gumbel-Softmax straight-through estimator in the backward pass, learned parameters are validated against the exact intended dynamics by construction.

The MNIST-trained gene regulatory network serves as a proof of principle for a new class of mechanistic machine learning. It illustrates that gradient descent can optimize large-scale stochastic reaction networks to implement non-trivial information processing. This positions stochastic biochemical kinetics not just as a system to be analyzed, but as a learnable substrate for computation, offering a rigorous alternative to black-box neural networks for modeling biological information processing. By transitioning from parameter-by-parameter sensitivity analysis to true massively-parallel backpropagation, our framework provides a concrete inverse-design workflow: specify a dynamic objective, represent the topology as a parameterized CTMC, and optimize hundreds of thousands of rate constants end-to-end while maintaining rigorous stochasticity. The ability to optimize such high-dimensional discrete systems supports a concrete inverse-design workflow, specify a dynamic objective, represent the topology as a parameterized CTMC, and optimize rate constants end-to-end while maintaining rigorous stochasticity. This capability complements emerging generative frameworks that automate the topological discovery of biomolecular networks, providing a rigorous numerical backbone for tuning the kinetics of generated candidates (*23*).

Our application to ion channel gating kinetics provides the critical stress test for this framework. In this regime, with only two channels and macroscopic observability of every transition, there is no quasi-continuous limit to exploit. The successful recovery of gating kinetics ($R^2 = 0.987$) confirms that our gradient estimator provides valid descent directions even when the system dynamics are dominated by individual discrete events. This result suggests that the method is robust enough for single-molecule biophysics and low-copy-number signaling contexts where continuum approximations fail entirely.

The mathematical framework developed here extends well beyond biochemistry. The Gillespie algorithm is a specific instance of the broader class of Kinetic Monte Carlo (KMC) methods, such as the Bortz-Kalos-Lebowitz (BKL) algorithm used in condensed matter physics. Our Gumbel-Softmax gradient estimator is isomorphic to these methods and is therefore immediately applicable to inverse problems in materials science, such as optimizing atomic potentials in crystal growth or tuning transition rates in defect migration, as well as stochastic epidemiological models and queueing networks. Any system governed by the master equation and simulated via competing Poisson processes can, in principle, be optimized using this differentiable framework.

The straight-through estimator is typically biased since the gradient it computes is a surrogate, not the true gradient of the expected objective. However, in the context of stochastic optimization, variance is often the more binding constraint. Unbiased likelihood-ratio estimators suffer from variance that explodes with trajectory length, rendering them useless for deep temporal models. Our results suggest that the low variance and strong correlation of the Gumbel-Softmax estimator outweigh the theoretical bias, providing a practical signal for optimization where unbiased alternatives diverge. Future theoretical work should further characterize the conditions under which this bias-variance trade-off is optimal.

Ultimately, our results provide a practical unification, exact discrete-event simulation becomes a backpropagation-compatible operator. By confining approximation to the gradient step and preserving exactness in the simulation, we remove the historical barrier between physical fidelity and scalable optimization. This enables gradient-based parameter inference and inverse design for stochastic systems at a scale previously reserved for differentiable surrogate models.

# References

1. M. B. Elowitz, A. J. Levine, E. D. Siggia, P. S. Swain, Stochastic gene expression in a single cell. *Science* **297**, 1183-1186 (2002).

2. H. H. McAdams, A. Arkin, Stochastic mechanisms in gene expression. *Proc Natl Acad Sci U S A* **94**, 814-819 (1997).

3. D. J. Wilkinson, *Stochastic Modelling for Systems Biology*. (Chapman and Hall/CRC, ed. 3rd, 2018).

4. C. Chang, M. Garcia-Alcala, L. Saiz, J. M. G. Vilar, P. Cluzel, Robustness of DNA looping across multiple cell divisions in individual bacteria. *Proceedings of the National Academy of Sciences* **119**, e2200061119 (2022).

5. I. Golding, J. Paulsson, S. M. Zawilski, E. C. Cox, Real-time kinetics of gene activity in individual bacteria. *Cell* **123**, 1025-1036 (2005).

6. J. M. G. Vilar, L. Saiz, Dynamics-informed deconvolutional neural networks for super-resolution identification of regime changes in epidemiological time series. *Science Advances* **9**, eadf0673 (2023).

7. R. Zandi, P. van der Schoot, D. Reguera, W. Kegel, H. Reiss, Classical nucleation theory of virus capsids. *Biophysical journal* **90**, 1939-1948 (2006).

8. J. M. G. Vilar, L. Saiz, Actionable Forecasting as a Determinant of Biological Adaptation. *Advanced Science* **12**, 2413153 (2025).

9. J. M. G. Vilar, L. Saiz, The unreasonable effectiveness of equilibrium gene regulation through the cell cycle. *Cell Systems* **15**, 639-648.e632 (2024).

10. A. Raj, A. Van Oudenaarden, Nature, nurture, or chance: stochastic gene expression and its consequences. *Cell* **135**, 216-226 (2008).

11. E. Korobkova, T. Emonet, J. M. G. Vilar, T. S. Shimizu, P. Cluzel, From molecular noise to behavioural variability in a single bacterium. *Nature* **428**, 574-578 (2004).

12. J. M. G. Vilar, R. V. Solé, Effects of noise in symmetric two-species competition. *Physical review letters* **80**, 4099 (1998).

13. D. T. Gillespie, Exact stochastic simulation of coupled chemical reactions. *The Journal of Physical Chemistry* **81**, 2340-2361 (1977).

14. A. B. Bortz, M. H. Kalos, J. L. Lebowitz, New Algorithm for Monte-Carlo Simulation of Ising Spin Systems. *Journal of Computational Physics* **17**, 10-18 (1975).

15. N. G. v. Kampen, *Stochastic processes in physics and chemistry*. North-Holland (Elsevier, Amsterdam ; Boston, ed. 3rd, 2007).

16. D. P. Kingma, J. Ba, Adam: A method for stochastic optimization. *International Conference on Learning Representations*. 2015.

17. F. Bach, Breaking the curse of dimensionality with convex neural networks. *Journal of Machine Learning Research* **18**, 1-53 (2017).

18. N. Altman, M. Krzywinski, The curse(s) of dimensionality. *Nature Methods* **15**, 399-400 (2018).

19. A. Golightly, D. J. Wilkinson, Bayesian parameter inference for stochastic biochemical network models using particle Markov chain Monte Carlo. *Interface Focus* **1**, 807-820 (2011).

20. T. Toni, D. Welch, N. Strelkowa, A. Ipsen, M. P. H. Stumpf, Approximate Bayesian computation scheme for parameter inference and model selection in dynamical systems. *Journal of The Royal Society Interface* **6**, 187-202 (2008).

21. R. J. Williams, Simple statistical gradient-following algorithms for connectionist reinforcement learning. *Machine Learning* **8**, 229-256 (1992).

22. Y. Bengio, N. Léonard, A. Courville, Estimating or propagating gradients through stochastic neurons for conditional computation. arXiv:1308.3432 (2013).

23. M. Filo, N. Rossi, Z. Fang, M. Khammash, GenAI-Net: A Generative AI Framework for Automated Biomolecular Network Design. *arXiv preprint arXiv:2601.17582*, (2026).

24. G. Tucker, A. Mnih, C. J. Maddison, J. Lawson, J. Sohl-Dickstein, Rebar: Low-variance, unbiased gradient estimates for discrete latent variable models. *Advances in Neural Information Processing Systems* **30**, (2017).

25. A. Gupta, M. Khammash, An efficient and unbiased method for sensitivity analysis of stochastic reaction networks. *Journal of The Royal Society Interface* **11**, (2014).

26. M. Rathinam, P. W. Sheppard, M. Khammash, Efficient computation of parameter sensitivities of discrete stochastic chemical reaction networks. *The Journal of Chemical Physics* **132**, (2010).

27. K. Rijal, P. Mehta, A differentiable Gillespie algorithm for simulating chemical kinetics, parameter estimation, and designing synthetic biological circuits. *eLife* **14**, RP103877 (2025).

28. E. Jang, S. Gu, B. Poole, Categorical reparameterization with Gumbel-Softmax. *International Conference on Learning Representations*. 2017.

29. C. J. Maddison, A. Mnih, Y. W. Teh, The Concrete distribution: A continuous relaxation of discrete random variables. *International Conference on Learning Representations*. 2017.

30. J. M. G. Vilar, H. Y. Kueh, N. Barkai, S. Leibler, Mechanisms of noise-resistance in genetic oscillators. *Proc Natl Acad Sci U S A* **99**, 5988-5992 (2002).

31. R. N. Gutenkunst, J. J. Waterfall, F. P. Casey, K. S. Brown, C. R. Myers, J. P. Sethna, Universally sloppy parameter sensitivities in systems biology models. *PLoS Comput Biol* **3**, 1871-1878 (2007).

32. Y. LeCun, L. Bottou, Y. Bengio, P. Haffner, Gradient-based learning applied to document recognition. *Proceedings of the IEEE* **86**, 2278-2324 (1998).

33. P. Y. Simard, D. Steinkraus, J. C. Platt, in *Seventh International Conference on Document Analysis and Recognition, 2003. Proceedings.* (IEEE Computer Society, 2003), vol. 3, pp. 958-958.

34. Z. Selimi, J.-S. Rougier, H. Abriel, J. P. Kucera, A detailed analysis of single-channel Nav1.5 recordings does not reveal any cooperative gating. *The Journal of Physiology* **601**, 3847-3868 (2023).

35. F. Mottes, Q.-Z. Zhu, M. P. Brenner, Gradient-based optimization of exact stochastic kinetic models. *arXiv preprint arXiv:2601.14183*, (2026).

36. Y. Cao, D. T. Gillespie, L. R. Petzold, Efficient step size selection for the tau-leaping simulation method. *J Chem Phys* **124**, 044109 (2006).

37. D. T. Gillespie, The chemical Langevin equation. *The Journal of Chemical Physics* **113**, 297-306 (2000).

38. C. Jia, R. Grima, Holimap: an accurate and efficient method for solving stochastic gene network dynamics. *Nature Communications* **15**, 6557 (2024).

39. E. J. Gumbel, *Statistical theory of extreme values and some practical applications: a series of lectures*. (US Government Printing Office, 1954), vol. 33.

40. B. T. Polyak, A. B. Juditsky, Acceleration of Stochastic Approximation by Averaging. *SIAM Journal on Control and Optimization* **30**, 838-855 (1992).

41. J. M. G. Vilar, L. Saiz, Systems biophysics of gene expression. *Biophys J* **104**, 2574-2585 (2013).

42. D. Colquhoun, On the stochastic properties of single ion channels. *Proceedings of the Royal Society of London. Series B. Biological Sciences* **211**, 205-235 (1981).

43. S. Matthew, F. Carter, J. Cooper, M. Dippel, E. Green, S. Hodges, M. Kidwell, D. Nickerson, B. Rumsey, J. Reeve, L. R. Petzold, K. R. Sanft, B. Drawert, GillesPy2: A Biochemical Modeling Framework for Simulation Driven Biological Discovery. *Lett Biomath* **10**, 87-103 (2023).

44. K. R. Sanft, S. Wu, M. Roh, J. Fu, R. K. Lim, L. R. Petzold, StochKit2: software for discrete stochastic simulation of biochemical systems with events. *Bioinformatics* **27**, 2457-2458 (2011).

## Acknowledgments

**Funding:** J.M.G.V. acknowledges support from Ministerio de Ciencia, Innovación y Universidades (MICIU/AEI/10.13039/501100011033/FEDER, UE) under Grant No. PID2024-160016NB-I00.

**Author contributions:** J.M.G.V. and L.S. designed research, performed research, analyzed data, and wrote the paper.

**Data availability:** Single-channel recordings are publicly available at Zenodo: https://zenodo.org/records/7817601. File used: HEK293_Cell01_40mV_100s_raw_DET_IDEl.xlsx.

**Code availability:** All the code used in this research will be made fully available on GitHub with the final version of this preprint.

## Figures

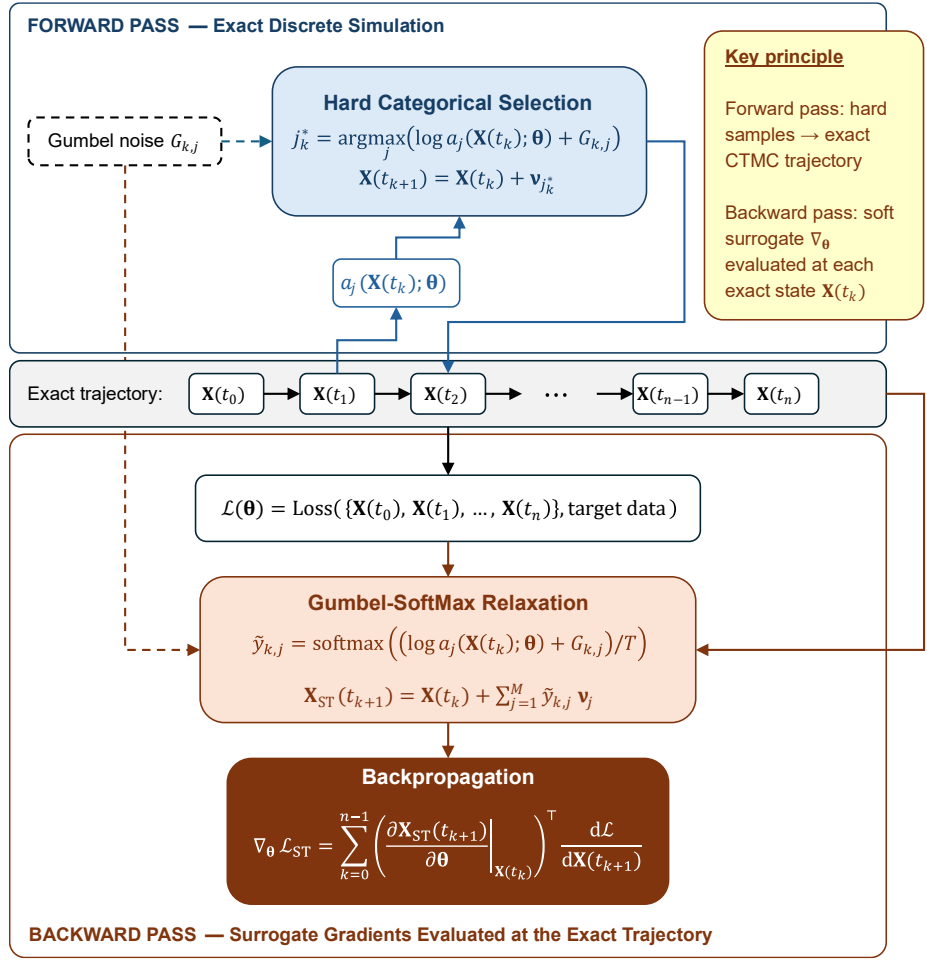

**Figure 1. Differentiable exact stochastic simulation via forward/backward decoupling.** The exact trajectory $\{\mathbf{X}(t_0), \mathbf{X}(t_1), \ldots, \mathbf{X}(t_n)\}$ (central row, $\mathbf{X}(t_k) \in \mathbb{Z}^d$) is the shared backbone for both computation passes. **Top (blue, forward pass):** At each step $k$, propensities $a_j(\mathbf{X}(t_k); \boldsymbol{\theta})$ are computed from the current discrete state, independent standard Gumbel variates $G_{k,j}$ are drawn, and the firing reaction is selected by hard argmax over perturbed log-propensities (Gumbel-Max trick). The state updates by the integer stoichiometric vector $\mathbf{v}_{j_k^*}$, preserving the full intrinsic noise of the continuous-time Markov chain. Waiting times between reactions are sampled by standard inverse-transform ($\tau = -\ln u / a_0$, $u \sim \text{Uniform}(0,1)$) and are omitted from the diagram for clarity. The loss $\mathcal{L}(\boldsymbol{\theta})$ is evaluated over the complete exact trajectory. **Bottom (red, backward pass):** The entire backward pass implements a Gumbel-Softmax straight-through estimator. At each step $k$, the same Gumbel variates $G_{k,j}$ from the forward pass enter a temperature-controlled softmax to yield differentiable soft reaction probabilities $\tilde{y}_{k,j} = \text{softmax}((\log a_j(\mathbf{X}(t_k); \boldsymbol{\theta}) + G_{k,j})/T)$, which define the surrogate state update $\mathbf{X}_{\text{ST}}(t_{k+1}) = \mathbf{X}(t_k) + \sum_{j=1}^{M} \tilde{y}_{k,j}\, \mathbf{v}_j$, where $T$ is the Gumbel-

Softmax temperature. The surrogate gradient $\nabla_{\boldsymbol{\theta}} \mathcal{L}_{\mathrm{ST}} = \sum_{k=0}^{n-1} (\partial \mathbf{X}_{\mathrm{ST}}(t_{k+1}) / \partial \boldsymbol{\theta}|_{\mathbf{X}(t_k)})^\top \mathrm{d}\mathcal{L} / \mathrm{d}\mathbf{X}(t_{k+1}) \in \mathbb{R}^p$ couples each per-step surrogate Jacobian with the backpropagated adjoint $\mathrm{d}\mathcal{L}/\mathrm{d}\mathbf{X}(t_{k+1}) \in \mathbb{R}^d$, which carries the accumulated sensitivity of $\mathcal{L}$ to $\mathbf{X}(t_{k+1})$ through all downstream surrogate transitions. The use of $\partial$ (partial) versus d (total) distinguishes local sensitivities, evaluated at the exact discrete state $\mathbf{X}(t_k)$ with the Gumbel draws held fixed, from total sensitivities propagated via the chain rule through the entire trajectory. When the loss depends on all states, the total derivative satisfies the recursion $\mathrm{d}\mathcal{L}/\mathrm{d}\mathbf{X}(t_n) = \partial \mathcal{L} / \partial \mathbf{X}(t_n)$ and, for $k = n - 1, \ldots, 0$, $\mathrm{d}\mathcal{L}/\mathrm{d}\mathbf{X}(t_k) = \partial \mathcal{L} / \partial \mathbf{X}(t_k) + (\partial \mathbf{X}_{\mathrm{ST}}(t_{k+1}) / \partial \mathbf{X}(t_k))^\top \mathrm{d}\mathcal{L}/\mathrm{d}\mathbf{X}(t_{k+1})$, where $\partial \mathcal{L} / \partial \mathbf{X}(t_k)$ is the direct dependence of the loss on state $\mathbf{X}(t_k)$. All surrogate derivatives are evaluated at the exact discrete states, not at relaxed or approximate states. This expression is exact for the surrogate computational graph; the only approximation is that the straight-through estimator is a biased gradient estimator for the true discrete process.

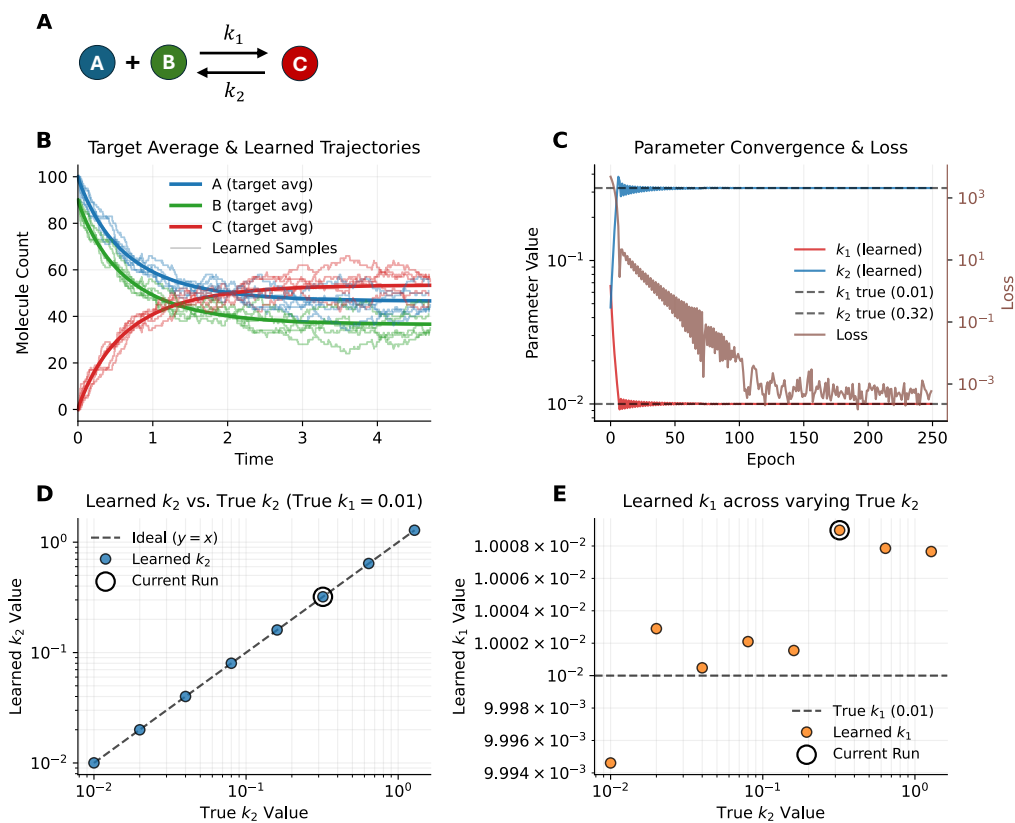

**Figure 2. Parameter inference for the reversible dimerization system.** (**A**) Reaction scheme. Two monomers A and B reversibly form a dimer C with forward rate constant $k_1$ and reverse rate constant $k_2$. This minimal system (2 parameters, 3 species, 2 reactions) serves as a validation benchmark. (**B**) Target and learned trajectories. Solid lines show ensemble-averaged target trajectories for species A (blue), B (green), and C (red) generated with ground truth parameters $k_1 = 0.01$ and $k_2 = 0.32$. Light traces show individual sample trajectories simulated with learned parameters. The learned dynamics closely match the target across the full time course, including transient and near-equilibrium regimes. (**C**) Parameter convergence and loss. Learned values of $k_1$ (red) and $k_2$ (green) converge to their true values (dashed lines) within approximately 50 epochs. The loss (gray, right axis) decreases by over three orders of magnitude during optimization. (**D**) Learned $k_2$ versus true $k_2$. Inference was repeated across a range of reverse rate constants $k_2 \in \{0.01, 0.02, 0.04, 0.08, 0.16, 0.32, 0.64, 1.28\}$ with $k_1 = 0.01$ held fixed. Each condition used 100,000 target trajectories and 100,000 model simulations per iteration (250 steps, 250 epochs). Learned values (green circles) align with the ideal $y = x$ relationship (dashed line) across two orders of magnitude, demonstrating accurate recovery regardless of kinetic regime. The circled point indicates the current run shown in panels B–C. (**E**) Learned $k_1$ across varying $k_2$. The inferred value of $k_1$ (green circles) remains within 0.1% of the true value (dashed line) across all tested $k_2$ values, confirming that parameter inference is robust and that $k_1$ and $k_2$ are independently identifiable. The circled point indicates the current run of panel B. Average MAPE across all conditions: 0.09% (95% CI: 0.06%–0.13%).

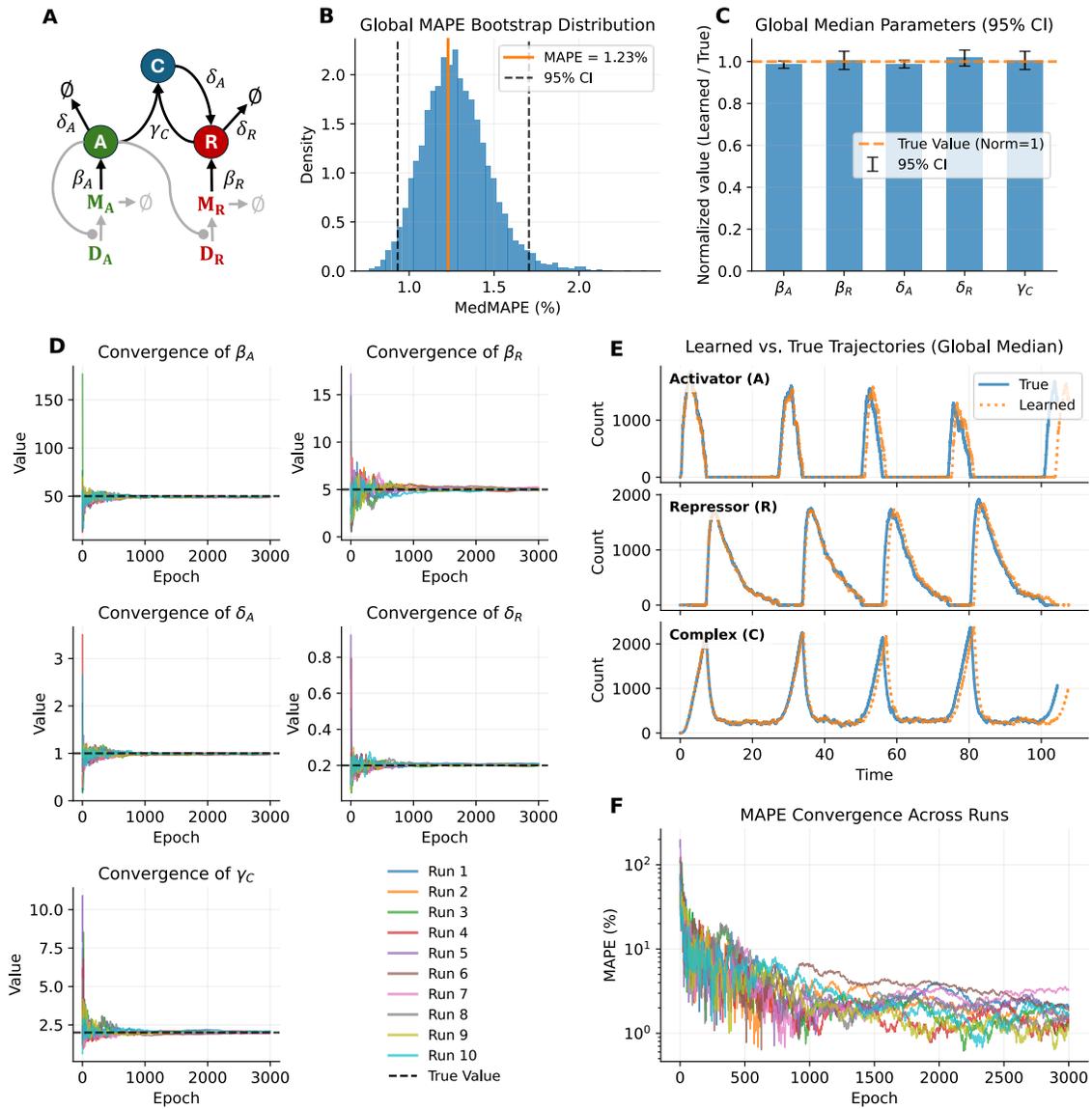

**Figure 3. Parameter inference for the genetic oscillator.** (**A**) Reaction network schematic. A gene regulatory network exhibiting sustained oscillations through coupled positive and negative feedback loops. The activator protein A promotes its own expression and activates repressor expression; the repressor R sequesters A through the formation of the C complex, creating delayed negative feedback. Five parameters controlling oscillatory dynamics ($\beta_A$, $\beta_R$, $\delta_A$, $\delta_R$, $\gamma_C$) were inferred from 9 species and 16 reactions. Black arrows indicate the reactions involving the inferred parameters. Grey lines and arrows indicate the schematics of the network not directly related to these parameters. (**B**) Bootstrap distribution of the global Mean Absolute Percentage Error (MAPE). Histogram shows the MAPE distribution of the estimated median parameters (median across runs of the last 25% percent values) computed from 10,000 block bootstrap resamples of the pooled parameter estimates across all runs. The observed MAPE of 1.23% (orange vertical line) falls near the center of the distribution. Black dashed vertical lines indicate the 95% bootstrap confidence interval. (**C**)

Global median parameter estimates with 95% confidence intervals. Bar heights show the ratio of learned to true parameter values for each of the five inferred parameters ($\beta_A, \beta_R, \delta_A, \delta_R, \gamma_C$). Error bars indicate 95% bootstrap confidence intervals. The orange dashed line at 1.0 represents perfect recovery. All parameters are recovered to within a few percent of their true values, with confidence intervals that include or nearly include the true value in all cases. (**D**) Individual parameter convergence trajectories over 3,000 training epochs, with different colors representing independent runs initialized from different random starting points. Black dashed lines indicate ground truth values. All parameters converge to near-true values within the first 1,000 epochs and remain stable thereafter. (**E**) Learned versus true trajectories. Representative single-trajectory simulations for the three key molecular species with ground truth parameters (solid blue) and learned parameters (dashed orange). (**F**) MAPE convergence across runs. MAPE (%) is plotted on a logarithmic scale as a function of training epoch for each run (same color coding as panel D). MAPE decreases from initial values of $10^2$–$10^3$% to approximately 1–5% by epoch 3,000, with most runs achieving errors near or below 2% in the final epochs.

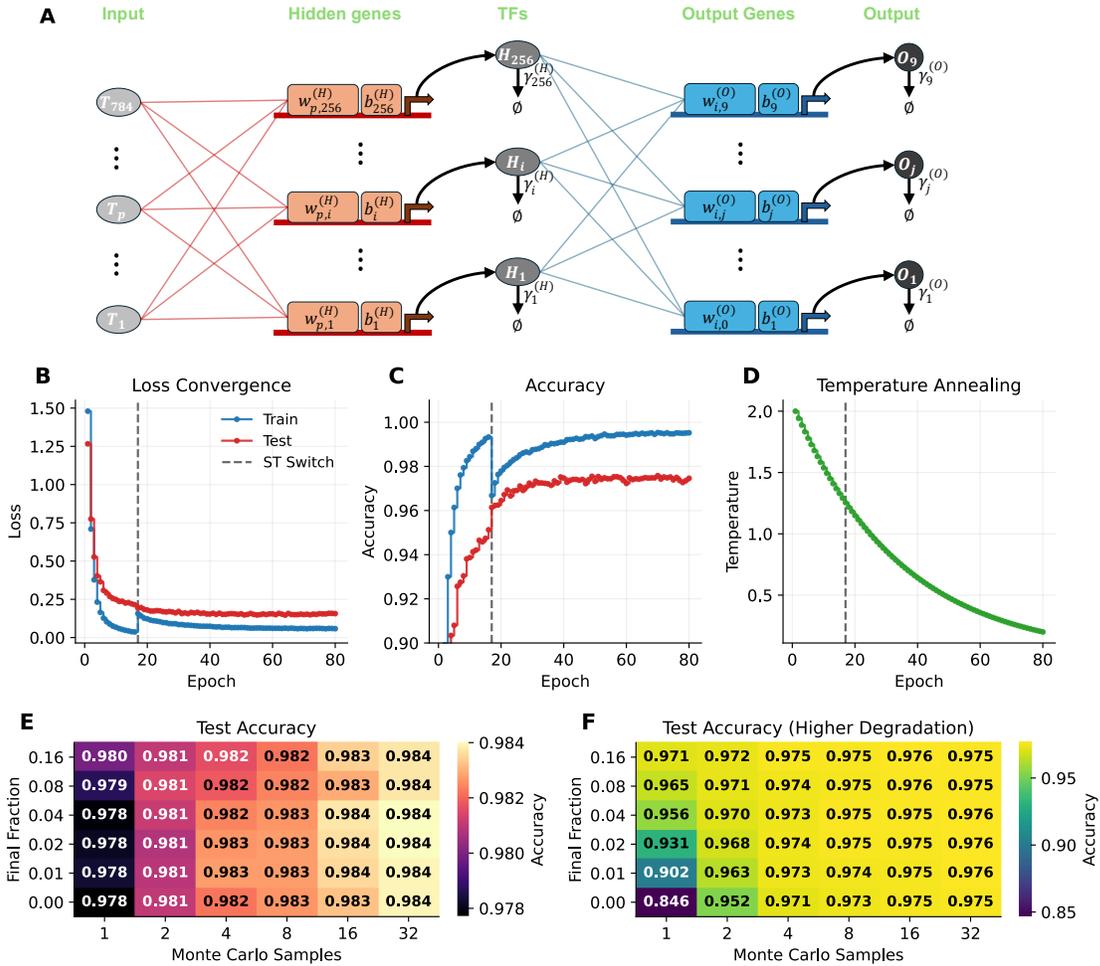

**Figure 4. Architecture, training, and evaluation of a gene regulatory network for MNIST digit classification.** (**A**) Network architecture. Input pixel intensities (28×28 = 784 values) define fixed transcription factor concentrations that regulate 256 hidden genes through a thermodynamic promoter model, where the probability of gene activation follows a sigmoid function $P(\text{ON}) = \sigma(E)$ with activation energy $E = b + \sum w \cdot x$. Hidden genes stochastically regulate 10 output genes corresponding to digit classes 0–9. Classification proceeds by simulating reaction steps and selecting the output gene with highest copy number. The network contains 203,796 trainable parameters (weights, biases, and degradation rates). (**B**) Loss convergence during training. Training loss (blue) and test loss (red) decrease over approximately 80 epochs. The vertical dashed red line indicates the epoch at which the straight-through (ST) estimator is switched on. (**C**) Classification accuracy during training. Training (blue) and test (red) accuracy increase over the same epoch range, reaching approximately 98% accuracy. (**D**) Temperature annealing schedule. The Gumbel-Softmax temperature is annealed during training, beginning at a high value to encourage exploration and decreasing to sharpen the categorical approximation for more precise gradient signals. (**E**) Test accuracy as a function of Monte Carlo averaging and temporal averaging. Heatmap shows classification accuracy on the MNIST test set as a function of the number of Monte Carlo (MC) samples (x-axis: 1–32) and the final fraction of simulation time steps over which output gene

expression is averaged for classification (y-axis: 0.00–0.16). A final fraction of 0.00 uses only the terminal time point; larger fractions average molecule counts over more time steps, reducing stochastic noise in the readout. (**F**) Test accuracy under higher degradation rate. The higher minimum degradation rate (switched from 0.25 to 1.0) produces lower baseline accuracy, but with larger gains from both MC and temporal averaging compared to panel E. In both settings, the largest accuracy gains come from MC averaging (moving right across columns), with diminishing returns beyond approximately 8 samples. These results demonstrate that the inherent stochasticity of the gene regulatory network can be leveraged through ensemble methods (e.g. populations of cells in biological systems) to achieve robust classification performance.

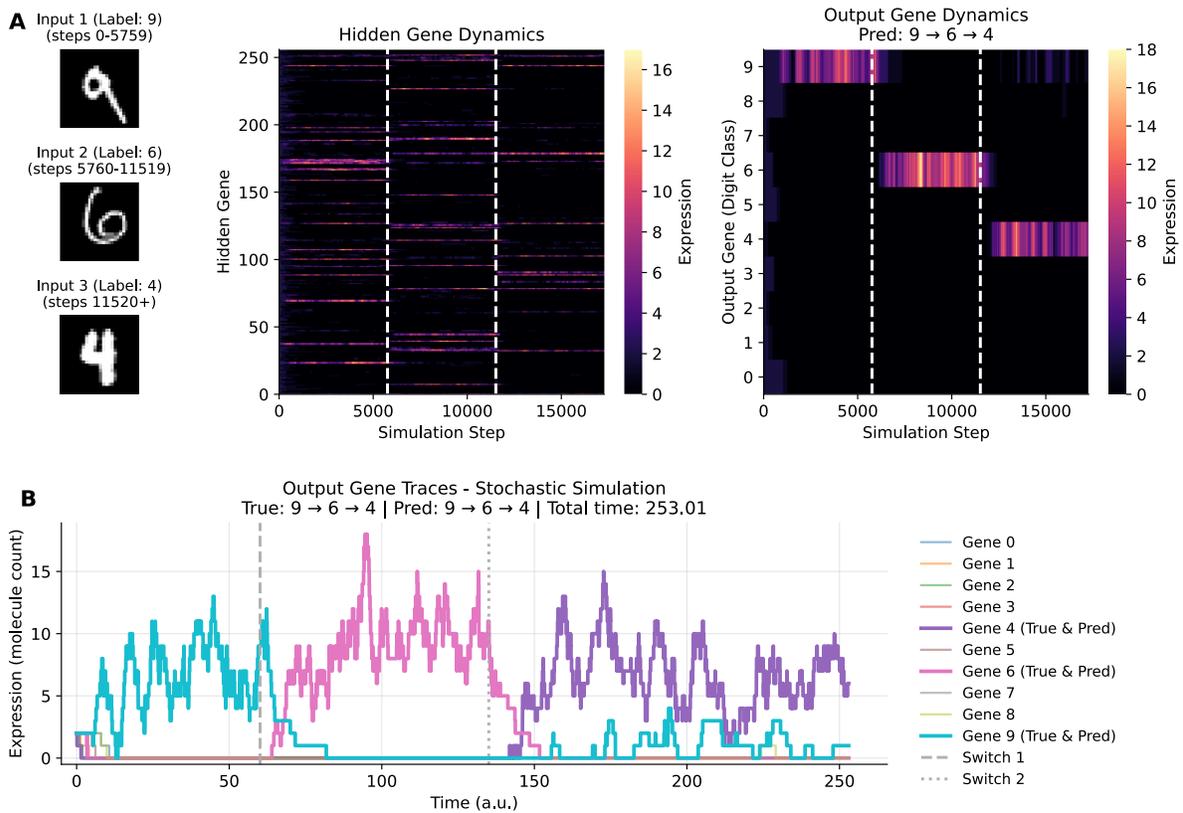

**Figure 5. Stochastic dynamics of the gene regulatory network during sequential digit classification.** (**A**) The network was presented with three consecutive MNIST images (digits 9, 6, and 4) to demonstrate real-time classification dynamics. (**Left**) Input images: digit "9" (simulation steps 0–5,759), digit "6" (steps 5,760–11,519), and digit "4" (steps 11,520+). (**Middle**) Hidden gene dynamics. Heatmap showing expression levels (molecule counts) for all 256 hidden genes over approximately 18,000 simulation steps. Dashed vertical lines indicate input image transitions. Different subsets of hidden genes activate in response to each digit, demonstrating that the hidden layer develops sparse digit-specific feature representations through training. (**Right**) Output gene dynamics. Heatmap showing expression levels for the 10 output genes corresponding to digit classes 0–9. During each input phase, the output gene matching the true digit label shows elevated expression while competing genes remain suppressed. The network correctly predicts all three digits (Pred: 9 → 6 → 4). (**B**) Output gene time traces. Continuous-time expression traces for all 10 output genes from a stochastic simulation, with genes corresponding to correct predictions highlighted: Gene 9 (cyan), Gene 6 (pink), and Gene 4 (purple). Dashed vertical lines mark input switching times. The stochastic winner-take-all dynamics resolve competition between output genes, with the correct class achieving dominant expression in each phase. True sequence: 9 → 6 → 4; Predicted sequence: 9 → 6 → 4.

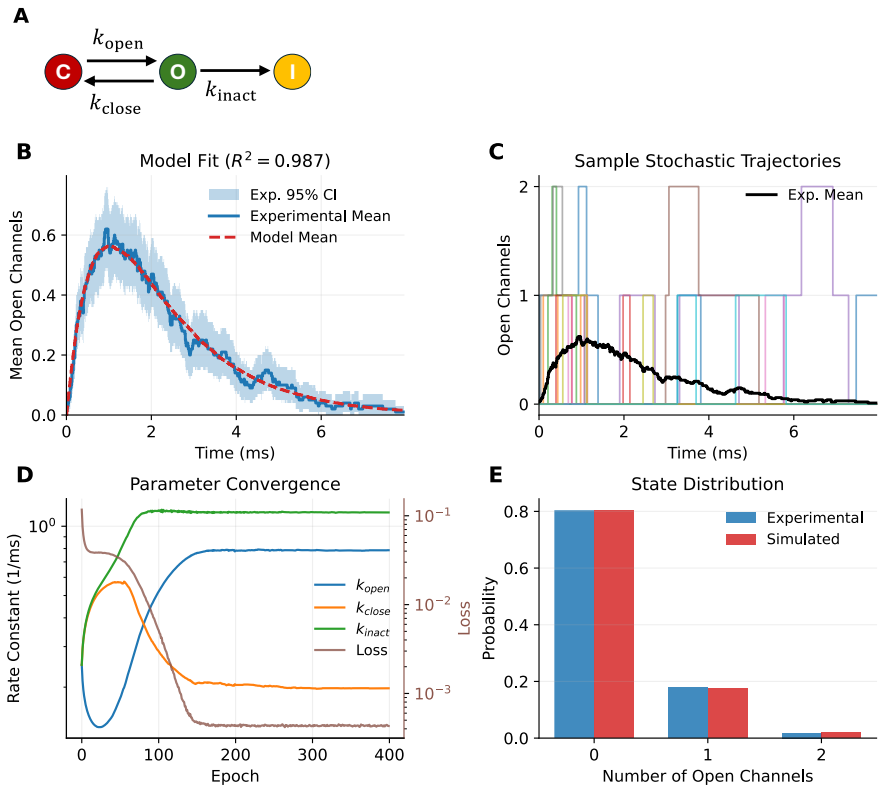

**Figure 6. Parameter inference for ion channel gating kinetics from patch-clamp recordings.** (**A**) Kinetic scheme. The three-state model consists of closed (C), open (O), and inactivated (I) conformations with first-order transitions governed by rate constants $k_{open}$, $k_{close}$, and $k_{inact}$. The inactivated state is absorbing, reflecting the absence of recovery on the experimental timescale. (**B**) Model fit to experimental data. Blue line shows the experimental ensemble mean open channel count (N = 100 sweeps) with shaded region indicating 95% bootstrap confidence interval. Red dashed line shows the model prediction using learned rate constants, evaluated from independent exact Gillespie simulations ($R^2 = 0.987$). The model captures both the rapid activation phase (peak at ~1 ms) and the slower inactivation decay. (**C**) Sample stochastic trajectories. Representative simulations (colored lines) using the learned parameters show discrete transitions between 0, 1, and 2 open channels. With only N = 2 channels, every stochastic event produces a macroscopic change in the observable. This extreme discreteness regime tests whether the method relies on quasi-continuous dynamics. Black line shows experimental mean for reference. (**D**) Parameter convergence during training. Rate constants evolve from initial guess toward stable values over 400 epochs. The inactivation rate $k_{inact}$ (green) converges fastest as it dominates long-time behavior, while $k_{open}$ (blue) and $k_{close}$ (orange) require additional iterations to resolve the transient dynamics. Loss (brown, right axis) decreases by over two orders of magnitude during training. (**E**) State distribution. Comparison of experimental (blue) and simulated (red) probability distributions over the number of open channels (0, 1, or 2). The close agreement confirms that the learned parameters reproduce the correct equilibrium occupancy of channel states. Data: HEK293 cells, +40 mV holding potential, 100 sweeps.

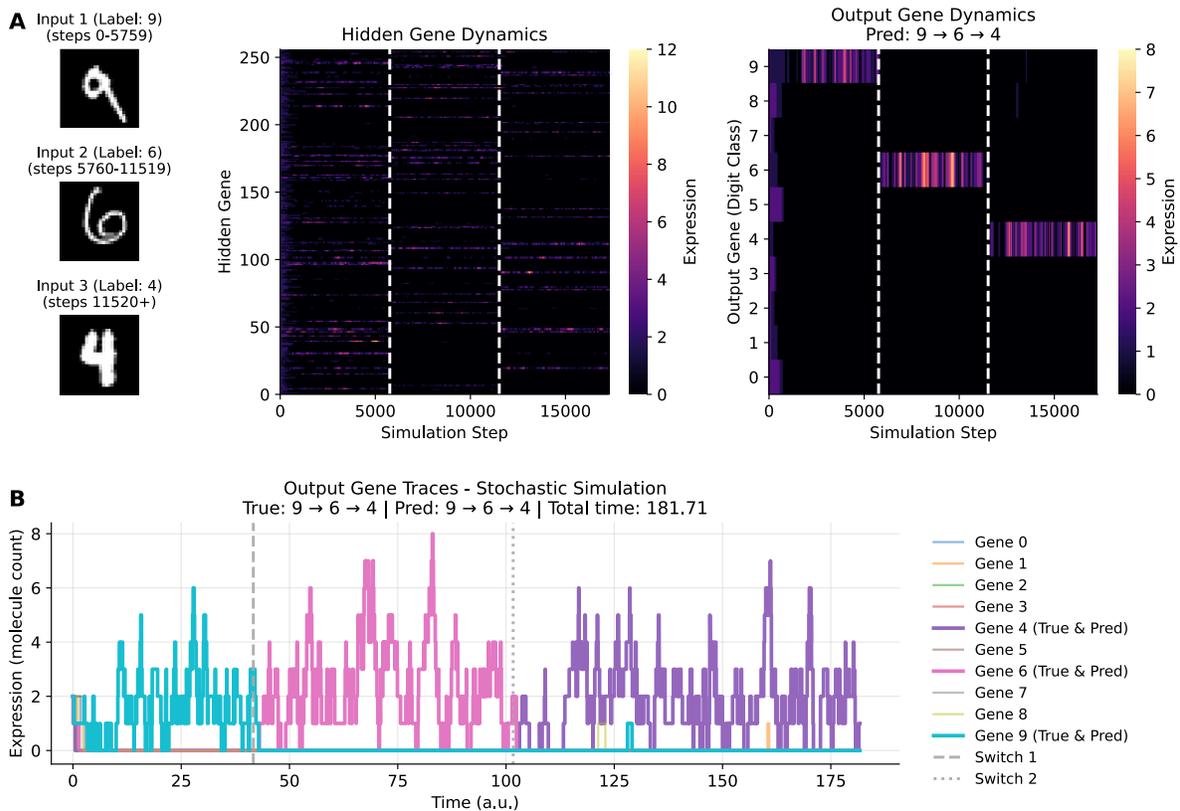

**Supplementary Figure 1. Stochastic dynamics of the gene regulatory network during sequential digit classification with a degradation rate lower bound of 1.0.** Same format, input sequence, and network architecture as Figure 5 (digits 9, 6, and 4), but using the model trained with a higher degradation rate (lower bound = 1.0; see Figure 4F). (**A**) (**Left**) Input images presented sequentially: digit "9" (simulation steps 0–5,759), digit "6" (steps 5,760–11,519), and digit "4" (steps 11,520+). (**Middle**) Hidden gene dynamics. Heatmap showing expression levels (molecule counts) for all 256 hidden genes over approximately 18,000 simulation steps. Dashed vertical lines indicate input image transitions. Despite the higher degradation rate, which increases molecular turnover and stochastic noise, distinct subsets of hidden genes activate in response to each digit. (**Right**) Output gene dynamics. Heatmap showing expression levels for the 10 output genes corresponding to digit classes 0–9. The network correctly classifies all three digits (Pred: 9 → 6 → 4). (**B**) Output gene time traces from the stochastic simulation. Continuous-time expression traces for all 10 output genes, with genes corresponding to correct predictions highlighted: Gene 9 (cyan), Gene 6 (pink), and Gene 4 (purple). Dashed vertical lines mark input switching times. The correct output gene achieves dominant expression during each input phase, demonstrating that the network maintains accurate classification even under elevated degradation, albeit with increased stochastic fluctuations in gene expression levels compared to the lower degradation rate regime (Figure 5). True sequence: 9 → 6 → 4; Predicted sequence: 9 → 6 → 4.

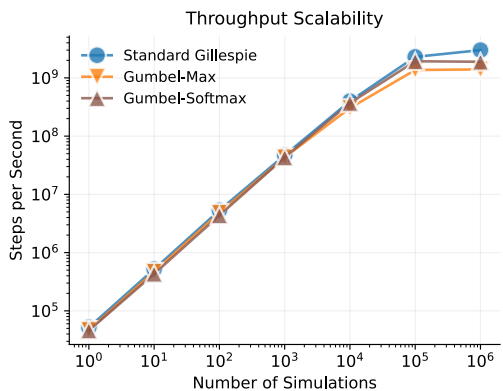

**Supplementary Figure 2. Computational throughput as a function of ensemble size.** Simulation throughput (steps per second) was measured for the oscillator system using three reaction selection methods: Standard Gillespie with categorical sampling (blue circles), Gumbel-Max with argmax selection (orange inverted triangles), and Gumbel-Softmax with temperature-controlled soft sampling (brown triangles). All methods show near-linear scaling with ensemble size from 1 to $10^5$ parallel simulations. At large ensemble sizes, throughput reaches approximately $10^9$ steps per second, with all three methods achieving comparable performance. The near-identical throughput across methods demonstrates that the differentiable framework incurs minimal computational overhead compared to non-differentiable implementations. Benchmarks were performed on an NVIDIA RTX 6000 Ada Generation GPU with TensorFlow 2.20.